# Projection-based Implicit Modeling Method (PIMM) for Functionally Graded Lattice Optimization

Hao Deng and Albert C. To[*]

Department of Mechanical Engineering and Materials Science, University of Pittsburgh, Pittsburgh, PA 15261

*Corresponding author. Email: albertto@pitt.edu

**Abstract**

This paper proposes a projection-based implicit modeling method (PIMM) for functionally graded lattice optimization, which does not require any homogenization techniques. In this method, a parametric projection function is proposed to link the implicit function of functionally graded lattice with the finite element background mesh. To reduce the number of design variables, the radial basis function (RBF) is utilized to interpolate the implicit design field. The triply periodic minimal surface (TPMS) lattice is employed to demonstrate the proposed method. Compared with conventional homogenization-based topology optimization, the proposed method can effectively resolve the stress-constrained lattice design; for example, sharp corners are removed from the initial design after optimization. Several two- and three-dimensional lattice design examples are presented to solve the compliance and stress-constrained problems, for which GPU-based computing is adopted to accelerate the finite element analysis. The proposed PIMM method is flexible and can potentially be extended to design graded irregular porous scaffold and non-periodic lattice infill designs.

## 1. Introduction

Topology optimization draws great attentions in the past several decades and becomes an important tool for structural and multidisciplinary optimization since the pioneering work of Bendsøe and Kikuchi [1]. Recently, several parametric design methods have been reported, such as the Moving Morphable Component (MMC) and Moving Morphable Voids (MMVs) approaches by Guo et al [2-10]. Some other methods such as discrete elements representation method are also proposed and achieve great success in geometry and complexity control [11-13]. Compared to the 0-1 optimization problem, lattice optimization based on conventional topology optimization experienced a surge in interest recently due to the rapid development of additive manufacturing (AM) techniques [14]. Based on homogenization theory, several effective topology optimization methods have been proposed in Refs [15-34].

Cellular materials or lattice structures have been utilized in numerous applications and are common in nature, such as bone, wood, sponge, etc. Recently, these porous materials are designed to achieve multi-functional material for weight reduction, energy absorption or heat transfer [35, 36]. Lattice structures made by metal are widely applied to product design in the field of orthopedic regenerative medicine [37]. These include, for example, design of bone scaffolds and implants to replicate the biomechanical properties of host bones. Porous metallic structure is an ideal candidate for repairing or replacing damaged bone because of its tunable mechanical properties. More importantly, porous metals can be designed to be open-celled to promote in-growth of bone tissue, which accelerates the osseointegration process. Conventional processes are difficult or impossible to fabricate porous media due to intricate internal architecture. Recent advancement in additive manufacturing (AM) enables fabrication of lattice structures which has significantly increased the demands for implants with customized mechanical performance [37]. In fact, open-celled lattice structures are preferred for additive manufacturing for many reasons including [23]: a) inherent porosity can minimize residual stress to reducing printed part distortion, b) reduce support materials due to self-supporting unit cells, and c) no enclosed voids so that the powder can be easily removed. Several approaches are proposed in recent years to generate lattice structures, such as generic ground truss structure approaches etc. [23, 38]. Triply periodic minimum surfaces (TPMS) is becoming a promising microstructure for designing scaffolds due to its extraordinary mechanical performance. TPMS is composed of three-dimensional continuous smooth surfaces, for which the average curvatures at each surface point is zero. These structures are called biomimetic structures, which is widely found in biological systems in nature, such as butterfly wings. Due to its particular geometric properties, these structures have the advantages of lightweight, high strength, and high specific surface area. Another merit of TPMS is that these structures can be defined using implicit equations. Besides, TPMS has already been proven to be a versatile source for biomorphic scaffold designs, and provided a viable and stable environment to replace damaged bones because of the smooth bending properties and optimized fluid permeability [39]. In this paper, we focus on the design of functionally graded lattice structure with TPMS unit cell due to its extraordinary mechanical properties.

The conventional computer-aided design (CAD) technique creates geometric objects using surfaces, which is an ideal solution for visualization and conventional subtractive manufacturing process, such as computer numerical control (CNC) cutting machine tools. However, this surface-based shape representation method is not an ideal for designs for additive manufacturing (AM). The AM technology builds an object in a layer-by-layer way, which can be applied to print extremely complex designs such as porous scaffolds or lattice structures. In general, geometric model for AM can be implemented using voxels, tetrahedra, parametric solids, or implicit field function [40] defined in three-dimensional space. However, using voxel points or a set of tetrahedra can be expensive in terms of storage space. Moreover, representation with voxels or tetrahedra provide only an approximation of the real object. For parametric representation, it is extremely tedious and difficult to design lattice structures using constructive solid geometry (CSG) [41]. Compared to the above methods, a ready-to-print geometric object is to describe a geometry as a 3D function $F(x, y, z)$, which directly informs the AM machine to determine whether the point $P(x, y, z)$ should be printed. Recent research has shown that geometry modeling using implicit functions are particularly suitable for modeling lattice and porous media. This AM-friendly modeling method has attracted great attention from academia and industry for AM design and has already been used to model any complex geometry in general. More importantly, this implicit field modeling method for AM has already been commercialized and achieve great success by a software company called nTopology, Inc. The main advantages of implicit modeling are as follows: i) An implicit geometry is directly defined in the physical space, which can directly provide precise information of objects to a 3D printer. ii) Implicit modeling is a lightweight geometric modeling technique without requirement of massive storage space. iii) Recent research has shown that implicit functions are particularly suitable for modeling microporous structures [40, 42]. Therefore, using implicit function to model porous media or lattice scaffold is a more advanced and feasible approach for future AM-oriented design.

The present work proposed a new frame work to combine implicit geometry modeling with projection-based method to achieve functionally graded lattice design. The method has flexibility to design irregular and complex scaffolds compared with homogenization-based framework, since material periodicity is not required for the proposed method. The paper is organized as follows. In Section 2, we describe projection-based implicit modeling method (PIMM) for functionally graded lattice design in details by using the Triply Periodic Minimal Surface (TPMS) structure as an example. Sections 3 presents formulation of two different optimization problems, and sensitivity analysis of PIMM is derived based on the chain rule. The optimized designs based on the PIMM method is presented in Section 5. The paper ends with conclusions in Section 6.

## 2. Implicit Modeling for Functionally Graded lattice

### 2.1 Generation of Functionally Graded Lattices based on Implicit Modeling

In general, lattice unit cells can be constructed using surface-based representation method. Using the TPMS as an example, the structure can be defined by a implicit function (i.e. $f(x,y,z) = t$), where $t$ is the parameter that governs the offset from the level sets, and $t$ can vary in design domain. There are several different types of TPMS as described in Ref. [23]. The four typical structures of TPMS [43] are shown in Fig. 1. The TPMS can be utilized to create lattice structures with unique mechanical characteristics. Furthermore, the lattice structures generated by TPMS have a higher surface-to-volume ratio compared with traditional strut-based lattice structures [44]. The gyroid is one of the most popular TPMS with robust mechanical performance. In this paper, we focus on the lattice design based on the gyroid minimum surface.

(a) G ('Gyroid')  (b) P ('Schwarz P')  (c) D ('Diamond')  (d) N ('Neovius')

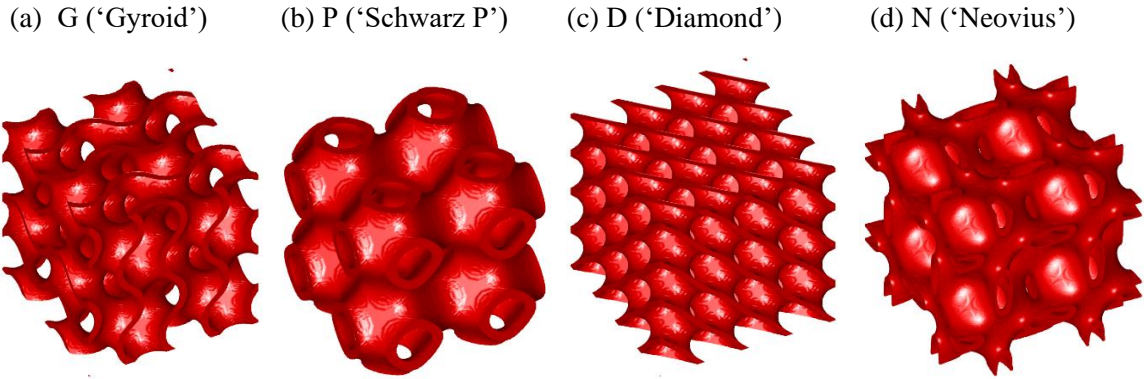

Figure 1. Triply periodic minimal surface (TPMS)

In general, TPMS can be constructed using inequality conditions expressed as [23]:

$$f(x,y,z)^2 \leq t \quad (t > 0) \tag{1}$$

The control equations for the gyroid surface is

$$f_G(x,y,z) = \sin(\lambda_x x) \cdot \cos(\lambda_y y) + \sin(\lambda_y y) \cdot \cos(\lambda_z z) + \sin(\lambda_z z) \cdot \cos(\lambda_x x) \tag{2}$$

where $\lambda_i (i = x, y, z)$ is the function periodicity, expressed as:

$$\lambda_i = \frac{2\pi}{L_i} \ (with\ i = x, y, z) \tag{3}$$

In the equations above, $L_i$ is the absolute dimension which defines the length of a unit cell. To design functionally graded lattice structures, the material grading in three-dimensional space can be realized by operating through 4D representation $(x, y, z, t)$, where the $t$ is an iso-value matrix in the $(x, y, z)$ space. Therefore, a functionally graded lattice can be represented in an implicit way as follows:

$$f(x,y,z)^2 \leq t(x,y,z) \tag{4}$$

In the equation above, the $t(x,y,z)$ controls the spatial variation of unit cell volume fraction in three-dimensional Cartesian space. Therefore, designing functionally graded (FG) TPMS lattice is equal to varying the variable $t(x,y,z)$ in the design domain. The design domain for $t(x,y,z)$ in Cartesian space can be discretized by voxels. The continuous function $t(x,y,z)$ in space can be reconstructed through field values at every voxel. An example of a uniform and linearly graded gyroid lattice is plotted in Fig. 2. In fact, the periodicity $\lambda_i$ can be also varied in three-dimensional Cartesian space. An example for gyroid lattice with different periodicity is presented in Fig. 3. In fact, the function periodicity $\lambda_i$ can also be a design variable in space for FG lattice design.

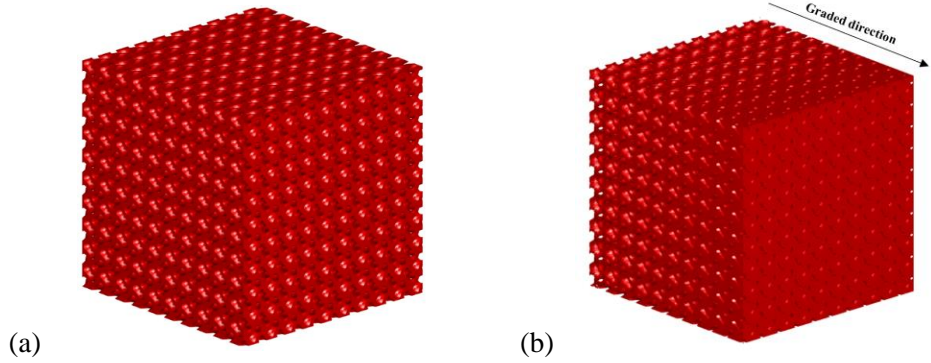

(a)          (b)

Figure 2. Lattice structures: (a) uniform lattice and (b) a linearly graded lattice

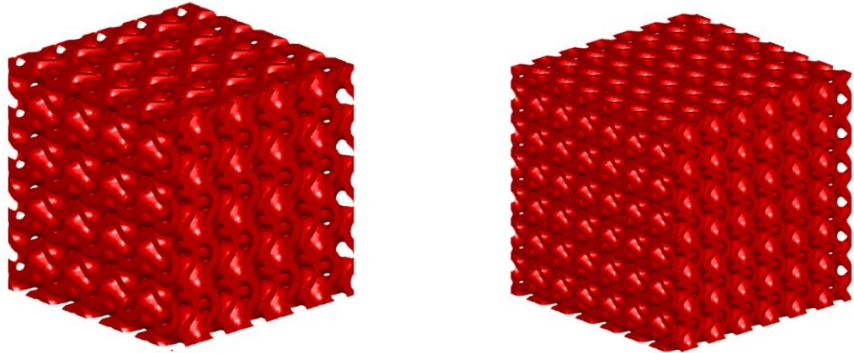

Figure 3. Gyroid lattice structure with two different periodicity

For conventional density-based topology optimization, design domain is discretized by finite element (FE) mesh, where each element works as a design variable. To connect the implicit field with density-based method, a Heaviside function-based projection method is implemented here to map the implicit field to the background FE mesh, which enables topology optimization algorithm to be performed on a fixed grid. An (approximate) Heaviside function can be defined as

$$H_{a,c}(x) = \frac{1}{1+e^{-a \cdot (x-c)}} \tag{5}$$

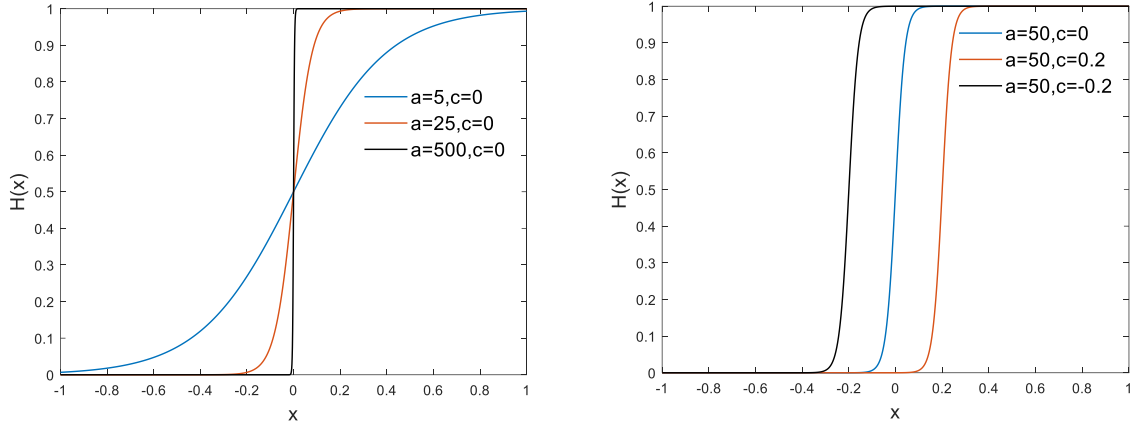

Figure 4. Heaviside Function

In the above equation, $a$ and $c$ are two parameters that could control the shape of Heaviside function. The shape of Heaviside function is plotted in Fig. 4 with different control parameters. A projection from the parametric design space $t(x, y, z)$ to density field $\rho$ can be expressed as:

$$\rho(x, y, z) = \mathrm{H}_{a_0, c_0}\big(t(x, y, z)\big) \cdot \mathrm{H}_{a_1, c_1}\big(t(x, y, z) - f(x, y, z)^2\big) \tag{6}$$

where $\mathrm{H}_{a_0, c_0}$ and $\mathrm{H}_{a_1, c_1}$ are Heaviside function defined in Eq. (5) with different control parameters. Note that the $t(x, y, z)$ is closely related to the volume fraction of unit cell. Thus, the first term is used to control the volume fraction of gyroid lattice. If the volume fraction at point $(x, y, z)$ is a small value, the first term will tend to zero so that the material at this point can be removed. The second term works as a projection to map the gyroid lattice to density field. Hence Eq. (6) is capable of mapping the design space $t(x, y, z)$ to density field $\rho(x, y, z)$. In practice, the design space $t(x, y, z)$ is a continuous differentiable function in design domain. To effective construct implicit field $t(x, y, z)$ in the entire design domain with a single globally continuous and differentiable function, the radial basis functions (RBFs) [45] is introduced here to model the implicit field $t(x, y, z)$. The RBFs are able to interpolate scattered data to generate smooth surface, and is an effective way to approximate complex function. Radial basis functions are radially symmetric functions centered at a specific point, called a RBF knot, which can be expressed as follows:

$$\varphi_i(x) = \varphi(\|x - x_i\|) \tag{7}$$

where $\|\cdot\|$ denotes the Euclidean norm and $x_i$ is the position of the knot. There are several possible radial basis functions, including thin-plate spline, Gaussians [46], etc. In this paper, The Gaussian function is chosen to work as the RBF kernel, where the explicit form of Gaussian function is expressed as follows:

$$\varphi(\|x - x_i\|) = e^{-\left(\frac{\|x - x_i\|}{\epsilon}\right)^2} \tag{8}$$

where $\epsilon$ is a parameter to control the shape of the Gaussian function. The implicit function $t(x)$ in the design domain can be interpolated via the RBF functions as:

$$t(x) = \sum_{i=1}^{N} \alpha_i \varphi_i(x) \tag{9}$$

where $\alpha_i$ is the expansion coefficient of the radial basis function positioned at the $i$th knot. The above equations can be rewritten as

$$t(x) = \phi^T(x)\alpha \qquad (10)$$

where

$$\phi(x) = [\varphi_1(x), \varphi_2(x), \cdots \varphi_N(x)]^T, \quad \alpha = [\alpha_1, \alpha_2, \cdots \alpha_N]^T \qquad (11)$$

A typical shape of Gaussian kernel in 2D space is plotted in Fig. 5.

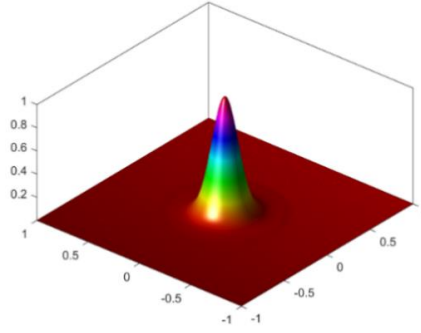

Figure 5. Shape of Gaussian kernel in 2D space

Using the RBFs to model the implicit function $t(x, y, z)$, we have

$$t(x, y, z) = \phi^T(x, y, z)\alpha_t \qquad (12)$$

where $\alpha_t$ is the design variable, which directly determines the implicit function $t(x, y, z)$. The relationship between RBF knot and density is illustrated in Fig. 6. While each density point is located at the center of each element of the FE mesh, the locations of RBF knots and density points can be independently chosen and do not need to coincide with each other. In our proposed scheme, the element densities are obtained through the projection function defined in Eq. (6), where the implicit field $t(x, y, z)$ is constructed using RBF knot through Eq. (12). Therefore, the density field $\rho(x, y, z)$ can be constructed via the following expression:

$$\rho(x, y, z) = H_{a_0, c_0}(\phi^T(x, y, z)\alpha_t) \cdot H_{a_1, c_1}(\phi^T(x, y, z)\alpha_t - f(x, y, z)^2) \qquad (13)$$

Note that the coordinate of $(x, y, z)$ should be normalized accordingly to the range of $[0,1]^3$ for optimization purpose.

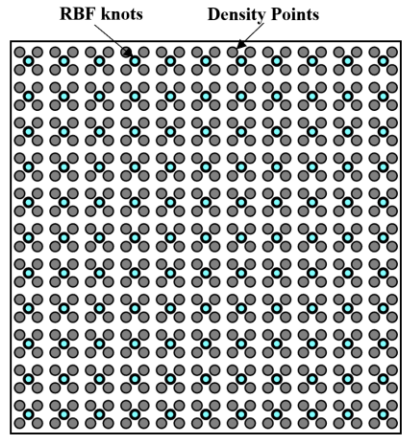

Figure 6. RBF knots and density points

## 2.2 Comparison with Homogenization-based Lattice Design

Homogenization-based functionally graded lattice design is a popular method in recent years [15, 47, 48]. The general procedure of this method is as follows: a) Compute the effective mechanical properties of unit cell by Asymptotic Homogenization (AH) method [49], b) conventional density-based TO method with effective material properties computed based on AH, and c) lattice reconstruction based on density optimized results, where the volume fraction of each unit cell is directly determined by material density distribution from density-based TO method. Although the homogenization-based method is widely accepted in industrial and academia, there still exist controversial and unresolved issues as follows:

1. Size effect of lattice structures. The AH method is effective only if the unit cell is sufficiently small in size compared to macrostructure. This condition is difficult to satisfy due to the current limitation of printing precision for metal components and computing power [50]; e.g., the element size in the homogenized FE model may be the same or slightly larger than lattice unit cell size.
2. Several advanced homogenization-based lattice design methods have been proposed in recent years [17, 18, 51], where the orientation and period of the unit cell are allowed to vary in space. However, whether the AH method is valid in such a situation is still controversial and requires further validation by theoretical and experimental means.
3. For stress-constrained lattice design problem, the sharp corners cannot be effectively removed and material tends to concentrate at the sharp corner due to shape-preserving characteristic of present homogenization-based method as described in Ref. [15]. However, removing the stress concentration region during lattice optimization is more reasonable and persuasive from the mechanics perspective.
4. Homogenization-based design method is not feasible for irregular porous scaffold designs [52], which is widely applied and preferred for tissue engineering. For example, Voronoi foam design [53], where there does not exist any periodicity in space, cannot be designed with homogenization-based method. However, because the irregular porous scaffold can be described by the implicit field [54], projection-based method proposed in this work can be readily applied in such situations.

The above issues for homogenization-based design method can be effectively resolved using the method proposed in this paper, which can be classified as a non-homogenization method for lattice design optimization based on implicit modeling.

## 3. Topology Optimization Formulation based on Implicit Modeling

### 3.1 Minimum Compliance

In this section, the implicit modeling method described in the section above is utilized to develop the TO formulation of compliance minimization [55]. The density field is controlled by RBF knots in the design domain. Hence, the TO will iteratively optimize functionally graded lattice through updating the RBF knots in the design domain until the design achieve the optimal stiffness. Here, the RBF knots are defined as the design variables for evolving the true density field in the design domain during the optimization. Thus, the optimization problem can be expressed as:

$$\begin{cases} \quad\quad\quad Find: \boldsymbol{\alpha_t} \\ Min: C(u, \Phi) = \frac{1}{2}\int_\Omega \boldsymbol{\varepsilon(u)}^T \boldsymbol{D}(\rho(\boldsymbol{\alpha_t}))\boldsymbol{\varepsilon(u)} d\Omega \\ \quad s.t: \frac{1}{|\Omega|}\int_\Omega \boldsymbol{\rho}(\boldsymbol{\alpha_t}) d\Omega - V_{prescribe} \leq 0 \end{cases} \quad (14)$$

where $C$ is the objective function defined by the structural compliance, $\alpha_t$ is the weight vector of the RBF knots in the design space, $\rho$ is the density distribution in the design domain $\Omega$, and $V_{prescribe}$ is the prescribed volume fraction. In the finite element model, $u$ is the unknown displacement field, $\varepsilon$ is the strain, and $D$ is the elastic tensor matrix.

### 3.2 Minimum Compliance with Stress Constraint

For the minimum compliance with stress constraint problem, the von Mises stress is always used for local stress measurement and as stress constraint in the optimization. However, constraining the local stress is numerically expensive in practice. Thus, a p-norm approach is implemented here to approximate the local stress constraint. Recently, several modified methods have been proposed to accurately control the local stress [56-64]. For simplicity, we apply a well-developed method to constrain the local von Mises stress as described in Ref. [65]. In this method, the p-norm measure $\sigma_{PN}$ is adopted to formulate the constraint. Thus, the problem in Sec. 3.1 can be reformulated as:

$$\begin{cases} \text{Find: } \alpha_t \\ \text{Min: } C(u,\rho) = \frac{1}{2}\int_\Omega \varepsilon(u)^T D(\rho(\alpha_t))\varepsilon(u)d\Omega \\ s.t.: \begin{cases} \frac{1}{|\Omega|}\int_\Omega \rho(\alpha_t)d\Omega - V_{prescribe} \leq 0 \\ \sigma_{PN} = \left(\sum_{e=1}^N \sigma_e^p\left(\rho(\alpha_t)\right)\right)^{\frac{1}{p}} \leq \overline{\sigma_{PN}} \end{cases} \end{cases} \quad (15)$$

where $p$ is the p-norm parameter, $\sigma_e$ is element von Mises stress, $\sigma_{PN}$ is p-norm measure, and $\overline{\sigma_{PN}}$ is the global stress limit. A good choice for $p$ can make the algorithm perform well and provide an adequate approximation of the maximum stress value. In this paper, $p = 10$ is applied in all stress-constrained numerical examples.

### 3.3 Design Sensitivity Analysis based on Chain Rule

To obtain the sensitivity of objective function with respect to weights of RBF knots, the chain rule is employed. The adjoint method [66] is applied to obtain the sensitivity with respect to the density field $\rho$:

$$\frac{\partial C}{\partial \rho} = \lambda^T \frac{\partial K}{\partial \rho} u \quad (16)$$

where $\lambda$ is the adjoint vector computed from the adjoint equation $K\lambda = -f$, and $K$ is the assembled stiffness matrix, see Ref. [55]. According to the chain rule, the sensitivity of objective $C$ with respect to design variables $\alpha_t$ can be expressed as:

$$\frac{\partial C}{\partial w} = \frac{\partial C}{\partial \rho} \cdot \frac{\partial \rho}{\partial \alpha_t} \quad (17)$$

The sensitivity of density field $\rho$ with respect to the RBF knots $\alpha_t$ can be readily obtained using symbolic differentiation system, which is available in a build-in module in MATLAB [67]. For sensitivity analysis of the p-norm stress, similar derivation can be achieved based on chain rule as follows:

$$\frac{\partial \sigma_{PN}}{\partial \alpha_t} = \frac{\partial \sigma_{PN}}{\partial \rho} \cdot \frac{\partial \rho}{\partial \alpha_t} \quad (18)$$

where the analytical sensitivity derivation based on the adjoint method of $\frac{\partial \sigma_{PN}}{\partial \rho}$ can be found in Ref. [68].

## 4. Numerical Examples and Discussion

In this section, several 2D and 3D numerical examples are demonstrated in details on designing functionally graded gyroid lattice structures. The classic MBB beam in two dimensions is first investigated to demonstrate the effective of the proposed implicit modeling method for lattice design. The parameters for all numerical examples are chosen as: $a_0 = 50$, $c_0 = 0.2$, $a_1 = 500$, $c_1 = 0$, $\epsilon = 0.1$. The Method of Moving Asymptotes (MMA) [69] is applied to solve optimization problem. The number of RBF knots in each direction are chosen based on our experience, less RBF knots result in simpler topology shape, while more RBF knots will inevitable increase computational cost. In this paper, ten RBF knots along each direction are implemented in all numerical examples. Note that for two-dimensional problem, the implicit function for describing the lattice structure $f_{2D}(x, y, z)$ is chosen as follows:

$$f_{2D}(x, y) = \sin(\lambda_x x) \cdot \cos(\lambda_y y) \tag{19}$$

**4.1 Compliance Optimization for MBB design**

The MBB-beam [70] is a popular test and benchmark problem in topology optimization. The symmetry is used for design, and the right half of the beam is modelled. The design of the MBB beam with the loading and boundary conditions is illustrated in Fig. 7. The design domain is uniformly meshed by 200×200 elements with unit length. The prescribed volume fraction is set as 30%. The elastic constants are chosen as follows: elastic modulus $E=1$ and Poisson's ratio $\mu=0.3$. The $10 \times 10$ uniformly distributed RBF knots are generated along two directions. The initial weights of RBF knots are chosen as 0.1. The periodicity parameters are selected as: $L_x = 0.1$, $L_y = 0.1$. The evolution of density field is presented in Fig. 8, and the optimized design is plotted in Fig. 9. The convergence history is plotted in Fig. 10. At the beginning of optimization, the design domain is infilled with uniformed lattice. In optimization, the density of every unit cell is changing gradually and materials will be selectively removed from the design domain as shown in Fig. 9. As shown in the optimal results, the method proposed in this paper is able to generate functionally graded lattice infill structures.

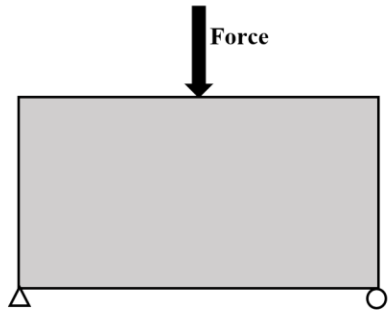

Figure 7. MBB beam example

Iteration: 1    Iteration: 10    Iteration: 20

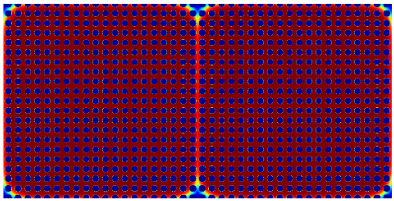 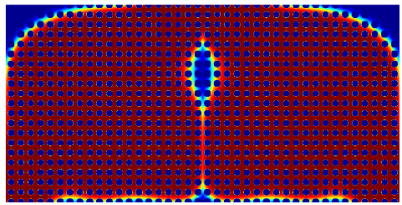 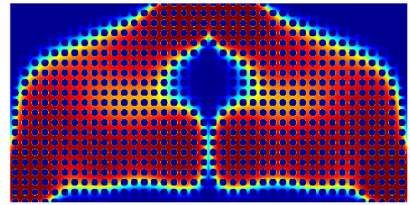

Iteration: 30   Iteration: 40   Iteration: 50

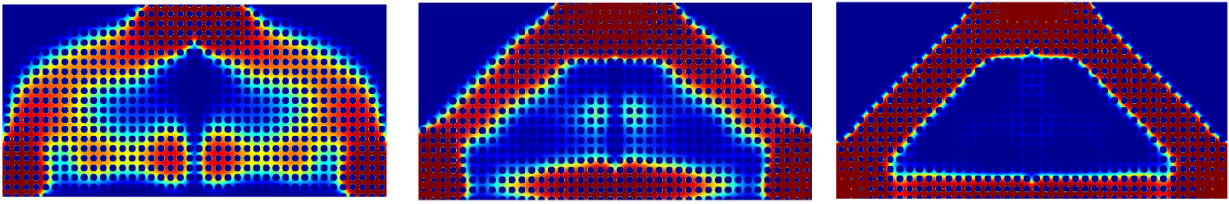

Figure 8. Evolution of density field

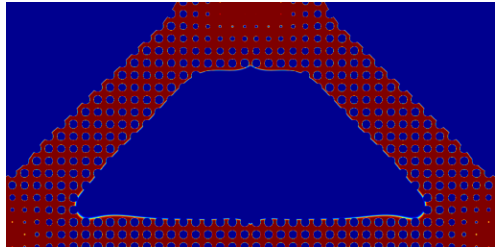

Figure 9. Optimal lattice infill design

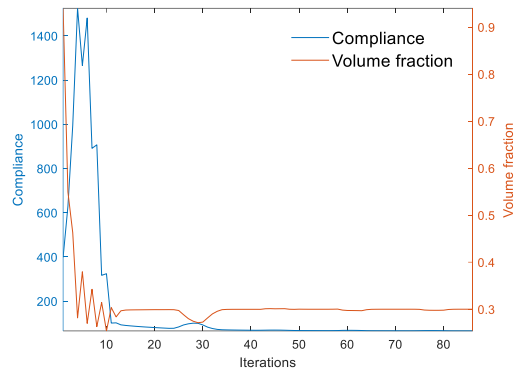

Figure 10. Convergence history

## 4.2 Stress Constrained Optimization for Two-Dimensional L-bracket Design

To further verify the effectiveness of the proposed implicit modeling method, the compliance minimization with stress constraint problem is considered in this section. The L-bracket is modeled by a $200 \times 200$ finite element mesh with $100 \times 100$ section (the density in this region is set to a small value $\rho = 1 \times 10^{-4}$ in FEM analysis) removed as shown in Fig. 11. The boundary condition and loading are demonstrated in Fig. 11. A vertical load $F = 4$ is applied uniformly on four nodes, and element size is unity in this numerical example. The topology optimization formulation for this example can be found in Eq. (15). The elastic constants are chosen as follows: Elastic modulus $E = 1$ and Poisson's ratio $\mu = 0.3$. The p-norm value for this numerical example is chosen as $p = 10$. The volume fraction is chosen as 0.3, and stress constraint in the p-norm is set to $\sigma_{pnorm} < 2$. The $10 \times 10$ uniformly distributed RBF knots are generated in the design domain as shown in Fig. 11. The initial weights of RBF knots are chosen as 0.1, and the periodicity parameters are selected as: $L_x = 0.05$, $L_y = 0.05$. Considering that the stress constraint optimization is highly nonlinear, a small moving limit of 0.1 in the MMA algorithm [71] is employed in the optimization. At the beginning of the optimization, the stress concentration occurs at the sharp corner, and the material in this area will be removed after optimization. The final optimal result is presented in Fig. 13. Note that round corners appear for the optimal lattice infill result to reduce stress concentration, and the outlier boundary of optimized material layout becomes smooth. The von Mises stress distribution for the final optimal design is presented in Fig. 13. The stress distribution of optimal design is uniform, and the maximum stress are approximately equal to 2. The convergence history is plotted in Fig. 14. Note that after optimization, the stress constraint can be satisfied, and the local oscillation of convergence curve can be observed in Fig. 14 due to the local stress singularity.

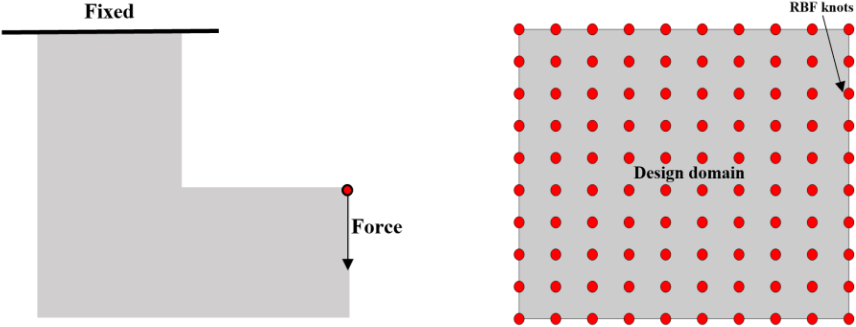

Figure 11. 2D L-bracket example

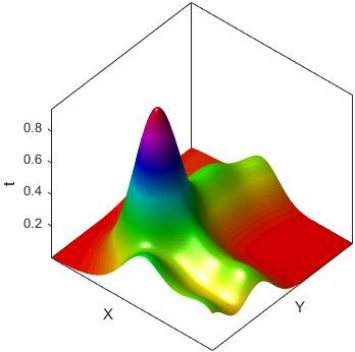

Figure 12. Distribution of implicit field $t(x,y)$.

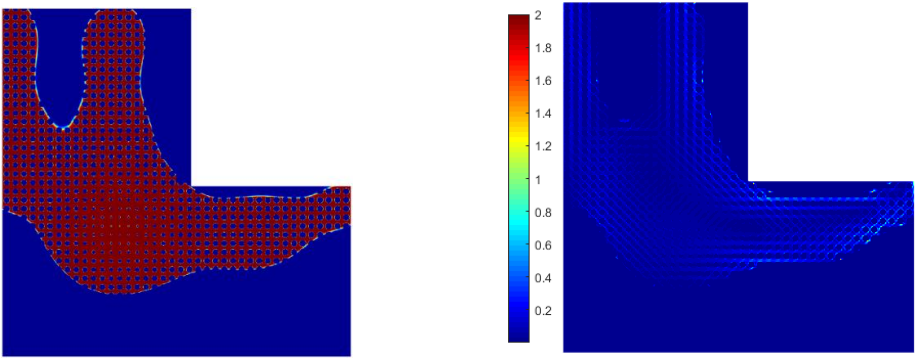

Figure 13. Optimized material layout and Von-mises stress distribution

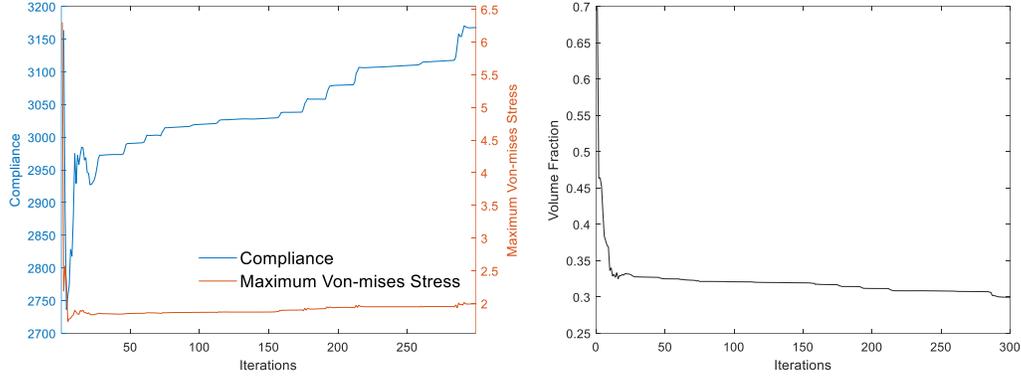

Figure 14. Convergence history

### 4.3 Compliance Optimization for Three-dimensional Cantilever Beam Design

In this section, a three-dimensional cantilever beam example is presented for compliance optimization. The cantilever beam is modeled by a $200 \times 100 \times 60$ hexahedral mesh, and the dimension of the design is demonstrated in Fig. 15. A uniform line force $F = 1$ is applied on the right-bottom of the rectangle domain. The $10 \times 10 \times 10$ uniformly distributed RBF knots in the design domain as shown in Fig. 15. The left side of rectangle is fixed. The elastic constants are chosen as follows: Elastic modulus $E = 1$ and Poisson's ratio $\mu = 0.3$. In this numerical example, the optimization progress is accelerated using GPU device based on the CUDA computing platform [72-74]. CUDA (Compute Unified Device Architecture) [75] is a parallel computing programming model developed by NVIDIA. CUDA provides a C language interface to the hardware, and a CUDA application including a host program on CPU, which can launch kernels written in CUDA to running on the parallel GPU device. The computational cost for large-scale FEM analysis mainly comes from two aspects, assembly of finite element and linear algebra solver. Reference [76] describes implementation of GPU acceleration of matrix assembly based on the efficient use of global, shared and local memory. For solving large-scale sparse linear systems, Multi-Grid Accelerated Linear Solvers (AMGX) [77] provides a simple and convenient way to accelerate linear solver on NIVIDIA GPUs with up to 10x acceleration. For three-dimensional examples, the FEA code is written and compiled with the NVIDIA CUDA Toolkit [75], and the numerical experiments are running on Linux OS with the NVIDIA Driver. The FEA solver employs a matrix-free PCG method with Jacobi preconditioner [77]. More details regarding implementation of the GPU acceleration of FEA can be found in Refs. [78, 79]. Note that only one GPU is used to accelerate the optimization progress in this paper. The initial weights of RBF knots are chosen as 0.1, and the periodicity parameters are selected as: $L_x = 0.1$, $L_y = 0.1$, and $L_z = 0.1$. The optimization converges after 30 iterations presented in Fig. 18, and the density evolution history is plotted in Fig. 17. The optimized lattice infill result is presented in Fig. 16.

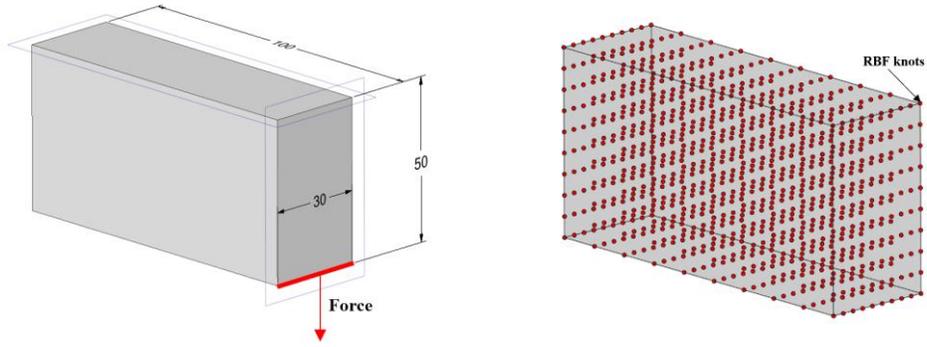

Figure 15. Three-dimensional Cantilever Beam

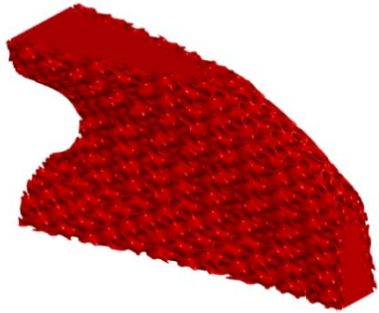

Figure 16. Optimal lattice infill design

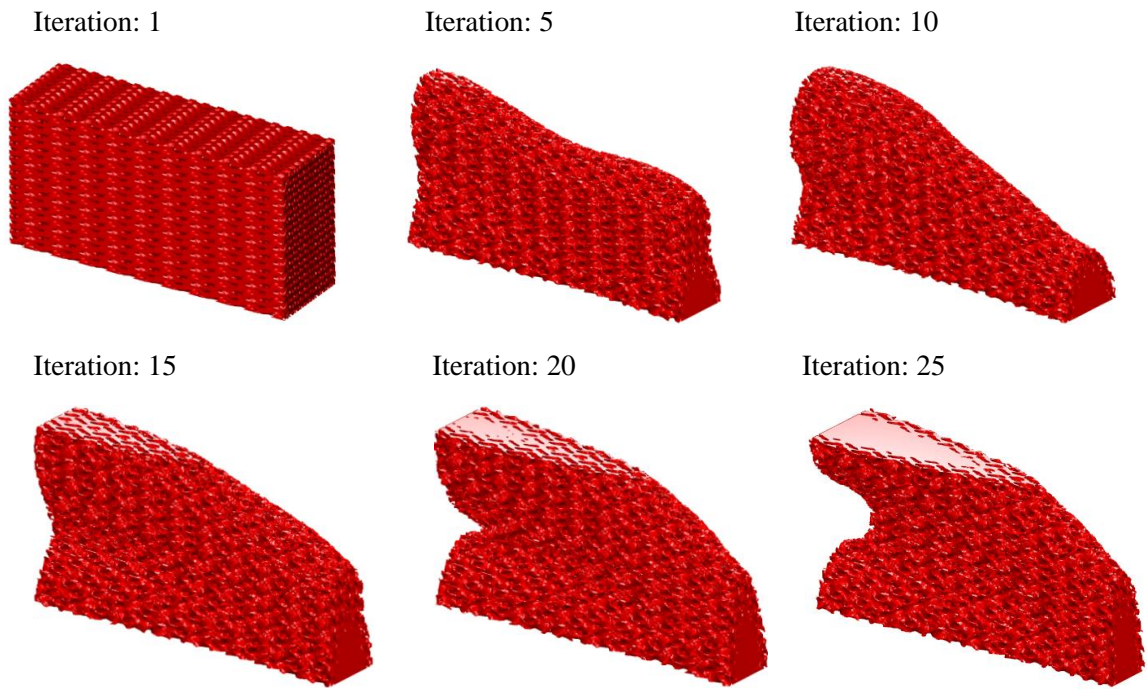

Figure 17. Evolution of density field

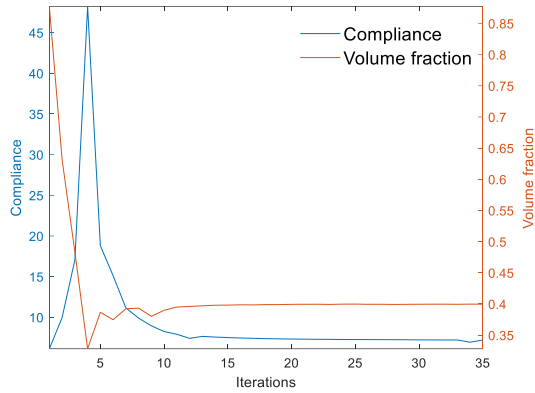

Figure 18. Convergence history

### 4.4 Compliance Optimization for 3D wheel Design

In this section, a 3D wheel Design example is presented for compliance optimization. The four corners are constrained by the planar joint with a point load $F = 1$ at the center as shown in Fig. 14. The 3D wheel is modeled by a $160 \times 160 \times 80$ hexahedral mesh, and the dimension of the design is demonstrated in Fig. 18. The $10 \times 10 \times 10$ uniformed distributed RBF knots in the design domain as demonstrated in Fig. 18. The initial weights of RBF knots are chosen as 0.1, and the periodicity parameters are selected as: $L_x = 0.1$, $L_y = 0.1$, and $L_z = 0.1$. The elastic constants are chosen as follows: Elastic modulus $E = 1$ and Poisson's ratio $\mu = 0.3$. GPU is applied in this example to accelerate the FEM analysis, where the basic setting for GPU acceleration is the same as the previous case. The initial design is shown in Fig. 20 (a). Actually, the method proposed in this work is able to produce shape-preserving results which are preferred for AM, because 0-1 topology optimization designs sometimes cannot be manufactured such as overhangs, and support structures [80] beneath them are needed. Furthermore, removing support structures is time-consuming and requires additional post-processing. To produce shape-preserving design, the value of parameter $c_0$ in Eq. (6) is set to be $c_0 = 0$. To make a comparison, two distinct optimization results with different values for parameter $c_0$ are demonstrated in Fig. 20 and Fig. 21 (same initial design), and convergence history for the two different designs is presented in Fig. 22. The optimized compliance values are close for these two different designs. As shown in Fig. 21, the optimized lattice structure is able to maintain the initial geometry configuration, where varied density lattice structures are generated, and materials tend to concentrate on four corners and loading point, and no block materials are removed. This shape preserving result is preferred for design of complex domain, where no overhang constraints are needed if the initial design domain is self-supported.

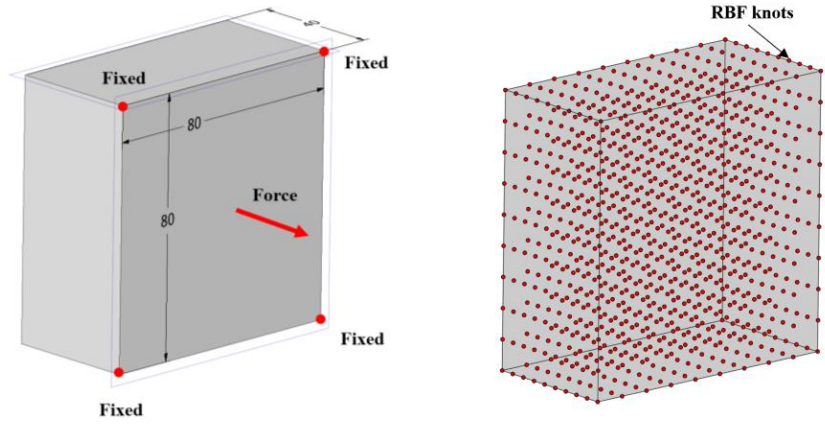

Figure 19. 3D wheel Design

(a) Initial design      (b) Front view      (c) Rear view

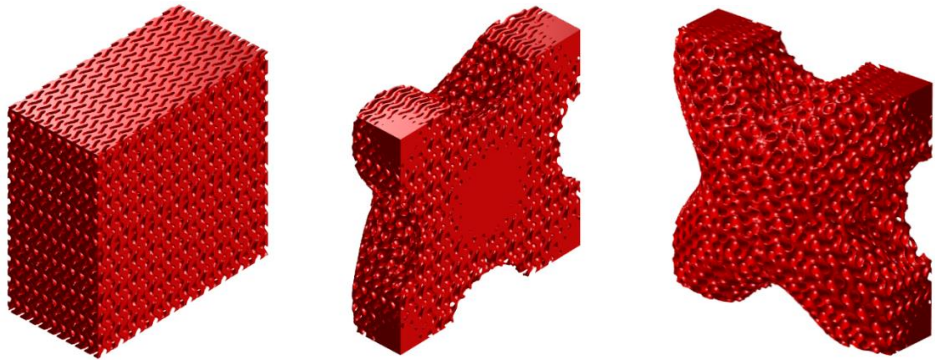

Figure 20. 3D wheel Design ($c_0 = 0.2$): (a) Initial design, (b) front view, and (c) rear view

(a) Front view      (b) Rear view

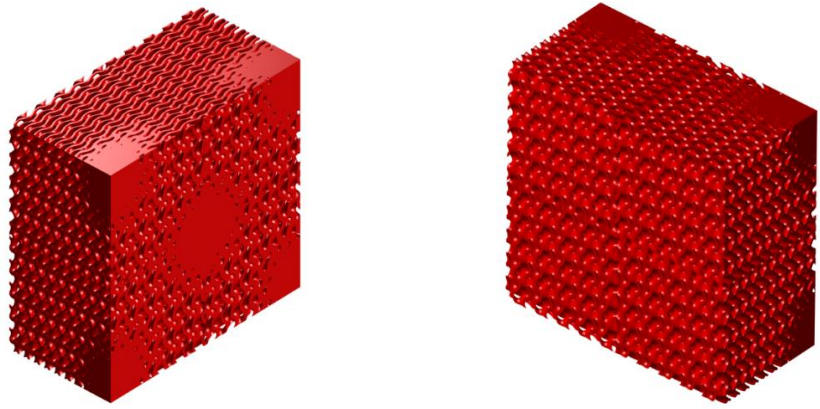

Figure 21. 3D wheel shape preserving design ($c_0 = 0$): (a) Front view and (b) rear view

(a) $c_0 = 0.2$  (b) $c_0 = 0$

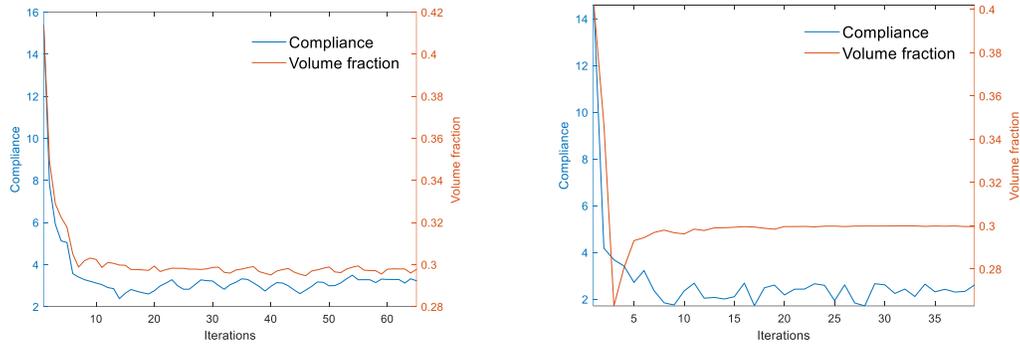

Figure 22. Convergence history for two design problems: (a) ($c_0 = 0.2$) and (b) ($c_0 = 0$)

### 4.5 Stress-Constrained Optimization for Three-dimensional L-bracket Design

To further test the proposed implicit field algorithm in a three-dimensional (3D) case, a 3D L-bracket example is plotted in Fig. 23. The dimensions of the L-bracket, and boundary and loading conditions are presented in Fig. 23. A distributed edge force $F = 1$ is applied to the FE model as shown in Fig. 23. The design domain is meshed with $200 \times 200 \times 60$ uniformly distributed trilinear hexahedral elements. The material properties are the same as the previous example. The p-norm stress constraint is set to be $\sigma_{pnorm} < 2$, and volume fraction constraint is chosen as 0.3. Due to the presence of a sharp corner in the initial design, the stress is expected to concentrate at the corner with a high value as shown in Fig. 25(a). The $10 \times 10 \times 10$ uniformly distributed RBF knots are generated in the design domain as demonstrated in Fig. 23, with the initial weight value 0.1, and the periodicity parameters are selected as: $L_x = 0.1$, $L_y = 0.1$, and $L_z = 0.1$. The small moving limit of MMA algorithm is chosen as 0.05. The computation is performed on the same CUDA configurations described above. The optimization converges after 70 iterations, and evolving density field during optimization is presented in Fig. 26. For homogenization-based stress constrained optimization of lattice structure, the optimized lattice structure is not able to removing sharp corner and the materials tend to concentrate at sharp corners to increase the yield strength of unit cells as described in Ref. [15]. Indeed, eliminating the sharp corners in optimization is a more reasonable solution from mechanics perspective, which can be successfully achieved by the proposed method. The optimized result is demonstrated in Fig. 24, where the material disappears near the region of sharp corners, and von Mises stress tends to be uniform for the optimal design. The von Mises stress distribution is plotted in Fig.

25(b), where the maximum von Mises stress of the initial design decreases from a value of 7 to 2 in the final design. The overall evolution of density distribution is presented in Fig. 26. The convergence history is plotted in Fig. 27, where the fluctuation happens at the beginning of the optimization progress. After 10 iterations, the optimization progress tends to become stable and both von Mises stress and volume fraction constraint are satisfied after optimization.

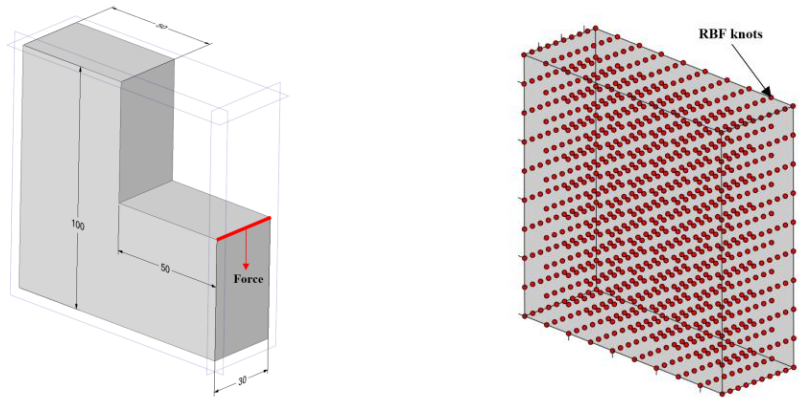

Figure 23. Three-dimensional L-bracket Design

(a) Front view    (b) Rear view

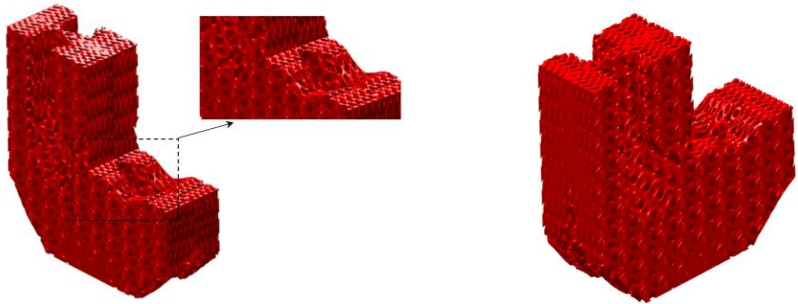

Figure 24. Optimized L-bracket Lattice Design: (a) Front view and (b) rear view

(a) Initial design    (b) Optimized lattice design

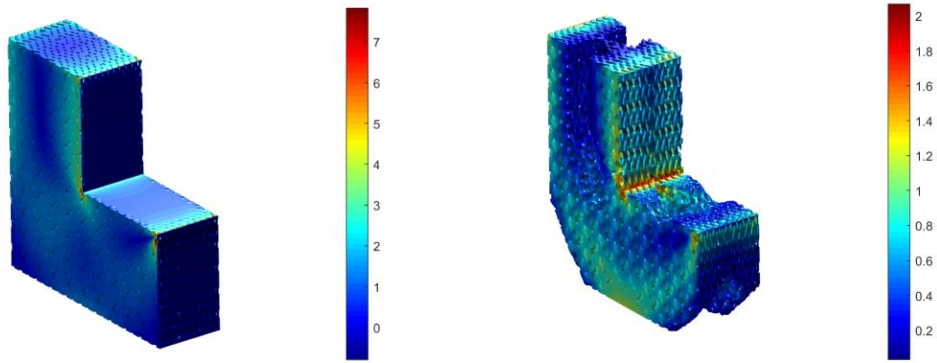

Figure 25. von Mises stress distribution in the design: (a) Initial design and (b) optimized design

Iteration: 1    Iteration: 5    Iteration: 15    Iteration: 25

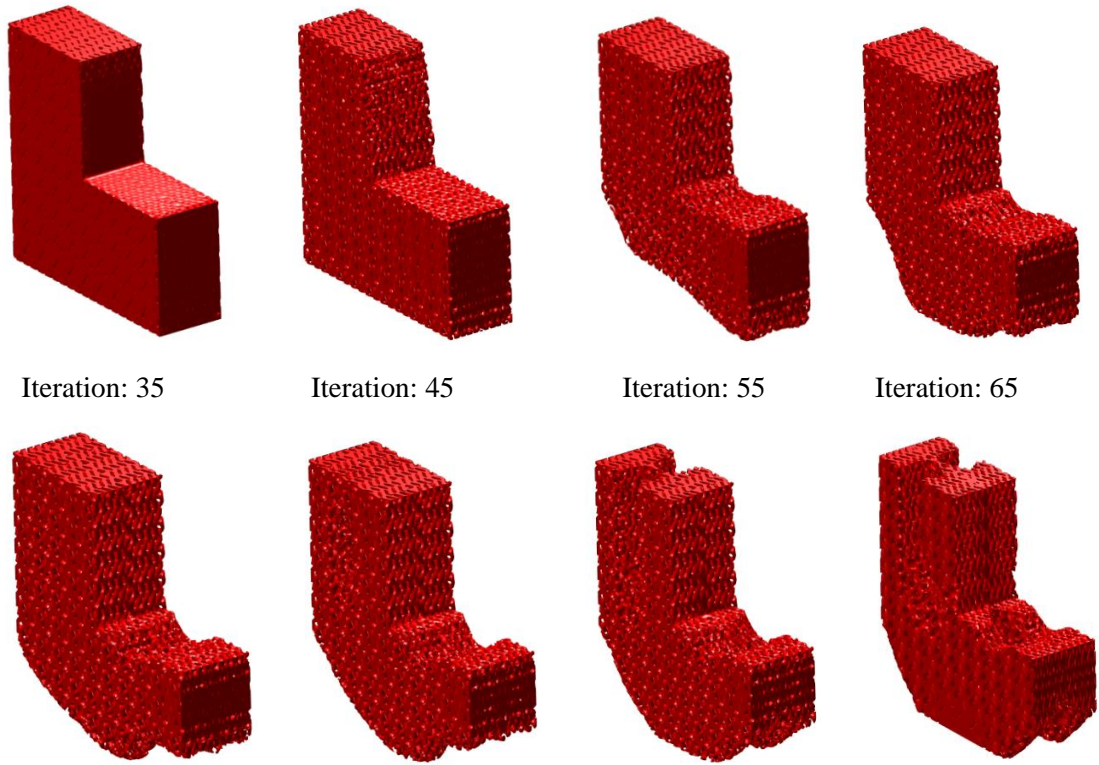

Figure 26. Evolution of density field

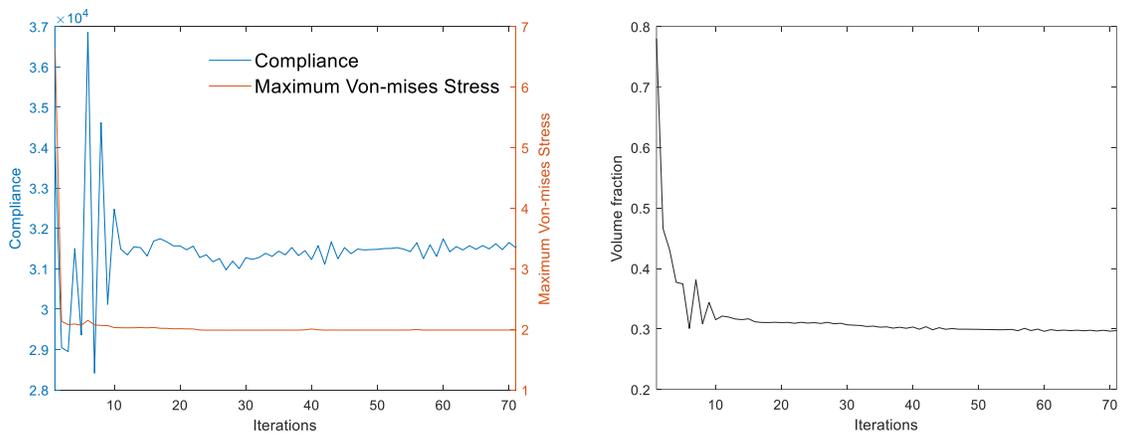

Figure 27. Convergence history

## 6. Conclusion

In this paper, a new projection-based algorithm based on implicit field for gyroid lattice design is proposed and demonstrated in details. The PIMM algorithm is able to design functionally graded lattice without the need for any homogenization. Thus, the lattice design based on this method is not limited to periodic structures, and can be extended to irregular porous scaffold designs. Besides relative density grading lattice design, the PIMM is also able to achieve cell size grading and cell type grading design [81], which is a natural extension of the present method and will be investigated in the future. Meanwhile, the unit cell size for lattice design can be large and not limited by size effects (homogenization necessary condition [82]), which is preferred for AM. Now that the geometry is defined by implicit function, the geometry information is far less than feature-based geometry modeling [83], which is sometimes extremely tedious for modeling porous media or lattice structures, and the data communication between implicit field with additive manufacturing systems is well-addressed by Ref. [40]. Furthermore, for stress constrained problems, homogenization-based method is hard to remove the stress concentration region, while the proposed implicit modeling method is able to resolve this problem for functionally graded lattice design, which is actually a combination of varied density lattice and material topology design.

## Acknowledgement

The financial support for this work from National Science Foundation (CMMI-1634261) is gratefully acknowledged.

## Reference

[1]     M. P. Bendsøe and N. Kikuchi, "Generating optimal topologies in structural design using a homogenization method," *Computer methods in applied mechanics engineering,* vol. 71, no. 2, pp. 197-224, 1988.
[2]     W. Zhang, D. Li, P. Kang, X. Guo, and S.-K. Youn, "Explicit topology optimization using IGA-based moving morphable void (MMV) approach," *Computer Methods in Applied Mechanics Engineering,* vol. 360, p. 112685, 2020.
[3]     R. Xue *et al.*, "Explicit structural topology optimization under finite deformation via Moving Morphable Void (MMV) approach," *Computer Methods in Applied Mechanics Engineering,* vol. 344, pp. 798-818, 2019.
[4]     C. Liu *et al.*, "An efficient moving morphable component (MMC)-based approach for multi-resolution topology optimization," *Structural Multidisciplinary Optimization,* vol. 58, no. 6, pp. 2455-2479, 2018.
[5]     W. Zhang, D. Li, J. Zhou, Z. Du, B. Li, and X. Guo, "A moving morphable void (MMV)-based explicit approach for topology optimization considering stress constraints," *Computer Methods in Applied Mechanics Engineering,* vol. 334, pp. 381-413, 2018.
[6]     W. Zhang, J. Zhou, Y. Zhu, and X. Guo, "Structural complexity control in topology optimization via moving morphable component (MMC) approach," *Structural Multidisciplinary Optimization,* vol. 56, no. 3, pp. 535-552, 2017.
[7]     W. Zhang *et al.*, "Explicit three dimensional topology optimization via Moving Morphable Void (MMV) approach," *Computer Methods in Applied Mechanics Engineering,* vol. 322, pp. 590-614, 2017.
[8]     W. Zhang, D. Li, J. Yuan, J. Song, and X. Guo, "A new three-dimensional topology optimization method based on moving morphable components (MMCs)," *Computational Mechanics,* vol. 59, no. 4, pp. 647-665, 2017.
[9]     X. Guo, W. Zhang, and W. Zhong, "Doing topology optimization explicitly and geometrically—a new moving morphable components based framework," *Journal of Applied Mechanics,* vol. 81, no. 8, 2014.
[10]    B. Zhu, Q. Chen, R. Wang, and X. Zhang, "Structural topology optimization using a moving morphable component-based method considering geometrical nonlinearity," *Journal of Mechanical Design,* vol. 140, no. 8, 2018.
[11]    J. Norato, B. Bell, and D. A. Tortorelli, "A geometry projection method for continuum-based topology optimization with discrete elements," *Computer Methods in Applied Mechanics and Engineering,* vol. 293, pp. 306-327, 2015.
[12]    S. Zhang, A. L. Gain, and J. A. Norato, "Stress-based topology optimization with discrete geometric components," *Computer Methods in Applied Mechanics and Engineering,* vol. 325, pp. 1-21, 2017.
[13]    S. Watts and D. A. Tortorelli, "A geometric projection method for designing three‐dimensional open lattices with inverse homogenization," *International Journal for Numerical Methods in Engineering,* vol. 112, no. 11, pp. 1564-1588, 2017.
[14]    W. E. Frazier, "Metal additive manufacturing: a review," *Journal of Materials Engineering and Performance,* vol. 23, no. 6, pp. 1917-1928, 2014.
[15]    L. Cheng, J. Bai, and A. C. To, "Functionally graded lattice structure topology optimization for the design of additive manufactured components with stress constraints," *Computer Methods in Applied Mechanics Engineering,* vol. 344, pp. 334-359, 2019.

[16] J. Groen, F. Stutz, N. Aage, J. A. Bærentzen, and O. Sigmund, "De-homogenization of optimal multi-scale 3D topologies," *arXiv preprint arXiv:.13002,* 2019.

[17] J. P. Groen, J. Wu, and O. Sigmund, "Homogenization-based stiffness optimization and projection of 2D coated structures with orthotropic infill," *Computer Methods in Applied Mechanics Engineering,* vol. 349, pp. 722-742, 2019.

[18] J. P. Groen and O. Sigmund, "Homogenization‐based topology optimization for high‐resolution manufacturable microstructures," *International Journal for Numerical Methods in Engineering,* vol. 113, no. 8, pp. 1148-1163, 2018.

[19] J. Wu, A. Clausen, and O. Sigmund, "Minimum compliance topology optimization of shell–infill composites for additive manufacturing," *Computer Methods in Applied Mechanics Engineering,* vol. 326, pp. 358-375, 2017.

[20] L. Cheng, P. Zhang, E. Biyikli, J. Bai, J. Robbins, and A. To, "Efficient design optimization of variable-density cellular structures for additive manufacturing: theory and experimental validation," *Rapid Prototyping Journal,* 2017.

[21] H. Zeinalabedini, Y. O. Yildiz, P. Zhang, K. Laux, M. Kirca, and A. C. To, "Homogenization of additive manufactured polymeric foams with spherical cells," *Additive Manufacturing,* vol. 12, pp. 274-281, 2016.

[22] P. Zhang *et al.*, "Efficient design-optimization of variable-density hexagonal cellular structure by additive manufacturing: theory and validation," *Journal of Manufacturing Science Engineering,* vol. 137, no. 2, 2015.

[23] A. Panesar, M. Abdi, D. Hickman, and I. Ashcroft, "Strategies for functionally graded lattice structures derived using topology optimisation for additive manufacturing," *Additive Manufacturing,* vol. 19, pp. 81-94, 2018.

[24] L. Xia and P. Breitkopf, "Multiscale structural topology optimization with an approximate constitutive model for local material microstructure," *Computer Methods in Applied Mechanics and Engineering,* vol. 286, pp. 147-167, 2015.

[25] L. Xia, *Multiscale structural topology optimization*. Elsevier, 2016.

[26] L. Xia and P. Breitkopf, "Recent advances on topology optimization of multiscale nonlinear structures," *Archives of Computational Methods in Engineering,* vol. 24, no. 2, pp. 227-249, 2017.

[27] Y. Zhu, S. Li, Z. Du, C. Liu, X. Guo, and W. Zhang, "A novel asymptotic-analysis-based homogenisation approach towards fast design of infill graded microstructures," *Journal of the Mechanics and Physics of Solids,* vol. 124, pp. 612-633, 2019.

[28] J. Gao, Z. Luo, H. Li, and L. Gao, "Topology optimization for multiscale design of porous composites with multi-domain microstructures," *Computer Methods in Applied Mechanics and Engineering,* vol. 344, pp. 451-476, 2019.

[29] J. Gao, Z. Luo, H. Li, P. Li, and L. Gao, "Dynamic multiscale topology optimization for multi-regional micro-structured cellular composites," *Composite Structures,* vol. 211, pp. 401-417, 2019.

[30] J. Fu, H. Li, L. Gao, and M. Xiao, "Design of shell-infill structures by a multiscale level set topology optimization method," *Computers & Structures,* vol. 212, pp. 162-172, 2019.

[31] R. Fleischhauer, T. Thomas, J. Kato, K. Terada, and M. Kaliske, "Finite thermo‐elastic decoupled two‐scale analysis," *International Journal for Numerical Methods in Engineering,* vol. 121, no. 3, pp. 355-392, 2020.

[32] J. Kato, D. Yachi, T. Kyoya, and K. Terada, "Micro‐macro concurrent topology optimization for nonlinear solids with a decoupling multiscale analysis," *International Journal for Numerical Methods in Engineering,* vol. 113, no. 8, pp. 1189-1213, 2018.

[33] S. Nishi, K. Terada, J. Kato, S. Nishiwaki, and K. Izui, "Two‐scale topology optimization for composite plates with in‐plane periodicity," *International Journal for Numerical Methods in Engineering,* vol. 113, no. 8, pp. 1164-1188, 2018.

[34] J. Kato, D. Yachi, K. Terada, and T. Kyoya, "Topology optimization of micro-structure for composites applying a decoupling multi-scale analysis," *Structural and Multidisciplinary Optimization,* vol. 49, no. 4, pp. 595-608, 2014.

[35] J. Brennan-Craddock, D. Brackett, R. Wildman, and R. Hague, "The design of impact absorbing structures for additive manufacture," in *Journal of Physics: Conference Series*, 2012, vol. 382, no. 1: IOP Publishing, p. 012042.

[36] I. Maskery *et al.*, "An investigation into reinforced and functionally graded lattice structures," *Journal of Cellular Plastics,* vol. 53, no. 2, pp. 151-165, 2017.

[37] X. Wang *et al.*, "Topological design and additive manufacturing of porous metals for bone scaffolds and orthopaedic implants: A review," *Biomaterials,* vol. 83, pp. 127-141, 2016.

[38] D. Brackett, I. Ashcroft, R. Wildman, and R. J. Hague, "An error diffusion based method to generate functionally graded cellular structures," *Computers Structures,* vol. 138, pp. 102-111, 2014.

[39]     S. C. Kapfer, S. T. Hyde, K. Mecke, C. H. Arns, and G. E. Schröder-Turk, "Minimal surface scaffold designs for tissue engineering," *Biomaterials,* vol. 32, no. 29, pp. 6875-6882, 2011.
[40]     Q. Li, Q. Hong, Q. Qi, X. Ma, X. Han, and J. Tian, "Towards additive manufacturing oriented geometric modeling using implicit functions," *Visual Computing for Industry, Biomedicine, Art,* vol. 1, no. 1, pp. 1-16, 2018.
[41]     A. Ricci, "A constructive geometry for computer graphics," *The Computer Journal,* vol. 16, no. 2, pp. 157-160, 1973.
[42]     D. J. Yoo, "Porous scaffold design using the distance field and triply periodic minimal surface models," *Biomaterials,* vol. 32, no. 31, pp. 7741-7754, 2011.
[43]     D. W. Abueidda, M. Elhebeary, C.-S. A. Shiang, S. Pang, R. K. A. Al-Rub, and I. M. Jasiuk, "Mechanical properties of 3D printed polymeric Gyroid cellular structures: Experimental and finite element study," *Materials Design,* vol. 165, p. 107597, 2019.
[44]     D.-J. Yoo, "Advanced porous scaffold design using multi-void triply periodic minimal surface models with high surface area to volume ratios," *International journal of precision engineering manufacturing,* vol. 15, no. 8, pp. 1657-1666, 2014.
[45]     E. Kansa, H. Power, G. Fasshauer, and L. Ling, "A volumetric integral radial basis function method for time-dependent partial differential equations. I. Formulation," *Engineering Analysis with Boundary Elements,* vol. 28, no. 10, pp. 1191-1206, 2004.
[46]     A. D. Cheng, M. Golberg, E. Kansa, and G. Zammito, "Exponential convergence and H‐c multiquadric collocation method for partial differential equations," *Numerical Methods for Partial Differential Equations: An International Journal,* vol. 19, no. 5, pp. 571-594, 2003.
[47]     D. Li, W. Liao, N. Dai, and Y. M. Xie, "Anisotropic design and optimization of conformal gradient lattice structures," *Computer-Aided Design,* vol. 119, p. 102787, 2020.
[48]     L. Cheng, J. Liu, and A. C. To, "Concurrent lattice infill with feature evolution optimization for additive manufactured heat conduction design," *Structural Multidisciplinary Optimization,* vol. 58, no. 2, pp. 511-535, 2018.
[49]     A. Bensoussan, J.-L. Lions, and G. Papanicolaou, *Asymptotic analysis for periodic structures*. American Mathematical Soc., 2011.
[50]     J. O. Milewski, *Additive manufacturing of metals: from fundamental technology to rocket nozzles, medical implants, and custom jewelry*. Springer, 2017.
[51]     D. Li, W. Liao, N. Dai, and Y. M. Xie, "Anisotropic design and optimization of conformal gradient lattice structures," vol. 119, p. 102787, 2020.
[52]     J. Feng, J. Fu, Z. Lin, C. Shang, and B. Li, "A review of the design methods of complex topology structures for 3D printing," *Visual Computing for Industry, Biomedicine, Art,* vol. 1, no. 1, p. 5, 2018.
[53]     G. Wang *et al.*, "Design and compressive behavior of controllable irregular porous scaffolds: based on voronoi-tessellation and for additive manufacturing," *ACS Biomaterials Science Engineering,* vol. 4, no. 2, pp. 719-727, 2018.
[54]     G. Wang *et al.*, "Design and compressive behavior of controllable irregular porous scaffolds: based on voronoi-tessellation and for additive manufacturing," vol. 4, no. 2, pp. 719-727, 2018.
[55]     E. Andreassen, A. Clausen, M. Schevenels, B. S. Lazarov, and O. Sigmund, "Efficient topology optimization in MATLAB using 88 lines of code," *Structural Multidisciplinary Optimization,* vol. 43, no. 1, pp. 1-16, 2011.
[56]     E. Lee, K. A. James, and J. R. Martins, "Stress-constrained topology optimization with design-dependent loading," *Structural Multidisciplinary Optimization,* vol. 46, no. 5, pp. 647-661, 2012.
[57]     R. Picelli, S. Townsend, C. Brampton, J. Norato, and H. Kim, "Stress-based shape and topology optimization with the level set method," *Computer methods in applied mechanics engineering,* vol. 329, pp. 1-23, 2018.
[58]     S. Zhang, A. L. Gain, and J. A. Norato, "Stress-based topology optimization with discrete geometric components," *Computer Methods in Applied Mechanics Engineering,* vol. 325, pp. 1-21, 2017.
[59]     C. Kiyono, S. Vatanabe, E. Silva, and J. Reddy, "A new multi-p-norm formulation approach for stress-based topology optimization design," *Composite Structures,* vol. 156, pp. 10-19, 2016.
[60]     H. Lian, A. N. Christiansen, D. A. Tortorelli, O. Sigmund, and N. Aage, "Combined shape and topology optimization for minimization of maximal von Mises stress," *Structural Multidisciplinary Optimization,* vol. 55, no. 5, pp. 1541-1557, 2017.
[61]     M. Zhou and O. Sigmund, "On fully stressed design and p-norm measures in structural optimization," *Structural Multidisciplinary Optimization,* vol. 56, no. 3, pp. 731-736, 2017.
[62]     S. Cai and W. Zhang, "Stress constrained topology optimization with free-form design domains," *Computer Methods in Applied Mechanics Engineering,* vol. 289, pp. 267-290, 2015.
[63]     L. Xia, L. Zhang, Q. Xia, and T. Shi, "Stress-based topology optimization using bi-directional evolutionary structural optimization method," *Computer Methods in Applied Mechanics Engineering,* vol. 333, pp. 356-370, 2018.

[64]	M. Y. Wang and L. Li, "Shape equilibrium constraint: a strategy for stress-constrained structural topology optimization," *Structural Multidisciplinary Optimization,* vol. 47, no. 3, pp. 335-352, 2013.
[65]	C. Le, J. Norato, T. Bruns, C. Ha, D. J. S. Tortorelli, and M. Optimization, "Stress-based topology optimization for continua," vol. 41, no. 4, pp. 605-620, 2010.
[66]	A. M. Bradley, "Pde-constrained optimization and the adjoint method," 2010.
[67]	E. Pärt-Enander, A. Sjöberg, B. Melin, and P. Isaksson, *The MATLAB handbook*. Addison-Wesley Harlow, 1996.
[68]	E. Holmberg, B. Torstenfelt, A. J. S. Klarbring, and M. Optimization, "Stress constrained topology optimization," vol. 48, no. 1, pp. 33-47, 2013.
[69]	K. Svanberg, "MMA and GCMMA-two methods for nonlinear optimization," vol. 1, pp. 1-15, 2007.
[70]	S. Rahmatalla and C. Swan, "A Q4/Q4 continuum structural topology optimization implementation," *Structural Multidisciplinary Optimization,* vol. 27, no. 1-2, pp. 130-135, 2004.
[71]	K. J. v. Svanberg, "MMA and GCMMA-two methods for nonlinear optimization," vol. 1, pp. 1-15, 2007.
[72]	G. Chakrabarti *et al.*, "CUDA: Compiling and optimizing for a GPU platform," *Procedia Computer Science,* vol. 9, pp. 1910-1919, 2012.
[73]	J. Thibault and I. Senocak, "CUDA implementation of a Navier-Stokes solver on multi-GPU desktop platforms for incompressible flows," in *47th AIAA aerospace sciences meeting including the new horizons forum and aerospace exposition*, 2009, p. 758.
[74]	D. Kirk, "NVIDIA CUDA software and GPU parallel computing architecture," in *ISMM*, 2007, vol. 7, pp. 103-104.
[75]	J. Sanders and E. Kandrot, *CUDA by example: an introduction to general-purpose GPU programming*. Addison-Wesley Professional, 2010.
[76]	C. Cecka, A. J. Lew, and E. Darve, "Assembly of finite element methods on graphics processors," *International journal for numerical methods in engineering,* vol. 85, no. 5, pp. 640-669, 2011.
[77]	M. Naumov *et al.*, "AmgX: A library for GPU accelerated algebraic multigrid and preconditioned iterative methods," vol. 37, no. 5, pp. S602-S626, 2015.
[78]	J. Martínez-Frutos, D. J. C. Herrero-Pérez, and Structures, "GPU acceleration for evolutionary topology optimization of continuum structures using isosurfaces," vol. 182, pp. 119-136, 2017.
[79]	S. Schmidt, V. J. C. Schulz, and V. i. Science, "A 2589 line topology optimization code written for the graphics card," vol. 14, no. 6, pp. 249-256, 2011.
[80]	D. Herzog, V. Seyda, E. Wycisk, and C. Emmelmann, "Additive manufacturing of metals," *Acta Materialia,* vol. 117, pp. 371-392, 2016.
[81]	I. Maskery, A. Aremu, L. Parry, R. Wildman, C. Tuck, and I. Ashcroft, "Effective design and simulation of surface-based lattice structures featuring volume fraction and cell type grading," *Materials Design,* vol. 155, pp. 220-232, 2018.
[82]	I. Goda, F. Dos Reis, and J.-F. Ganghoffer, "Limit analysis of lattices based on the asymptotic homogenization method and prediction of size effects in bone plastic collapse," in *Generalized Continua as Models for Classical and Advanced Materials*: Springer, 2016, pp. 179-211.
[83]	O. W. Salomons, F. J. van Houten, and H. Kals, "Review of research in feature-based design," *Journal of manufacturing systems,* vol. 12, no. 2, pp. 113-132, 1993.